\documentclass[fleqn,10pt]{wlscirep}
\usepackage[utf8]{inputenc}
\usepackage[T1]{fontenc}
\usepackage{multirow}
\usepackage{amsmath}
\usepackage{amssymb}

\title{Robust physics discovery via supervised and unsupervised pattern recognition using the Euler characteristic}

\author[1]{Zhiming Zhang}
\author[2]{Nan Xu}
\author[3*]{Yongming Liu}
\affil[1,2,3]{Arizona State University, School for Engineering of Matter, Transport and Energy, Tempe, 85281, USA}

\affil[*]{yongming.liu@asu.edu}

\begin{abstract}
Machine learning approaches have been widely used for discovering the underlying physics of dynamical systems from measured data. Existing approaches, however, still lack robustness, especially when the measured data contain a large level of noise. The lack of robustness is mainly attributed to the insufficient representativeness of used features. As a result, the intrinsic mechanism governing the observed system cannot be accurately identified. In this study, we use an efficient topological descriptor for complex data, i.e., the Euler characteristics (ECs), as features to characterize the spatiotemporal data collected from dynamical systems and discover the underlying physics. Unsupervised manifold learning and supervised classification results show that EC can be used to efficiently distinguish systems with different while similar governing models. We also demonstrate that the machine learning approaches using EC can improve the confidence level of sparse regression methods of physics discovery.
\end{abstract}
\begin{document}

\flushbottom
\maketitle

\thispagestyle{empty}

\section*{Introduction}
Scientific discovery has attracted increasing attention over the past years due to the boosting interest in exploring the new world. Data-driven approaches (machine learning, etc.) recently became possible and popular owing to revolutionary advances in sensing technologies and computational powers. Among all methods investigated for discovering scientific laws, sparse regression \cite{rudy2017data} gains the most attention in extracting scientific laws (expressed as ODEs or PDEs) from measured/simulated data. Sparse regression methods inherently promote the model parsimony which is widely recognized as an intrinsic property of the governing model for a real physical system. Moreover, techniques including genetic algorithm \cite{xu2020dlga}, reinforcement learning \cite{bassenne2019computational}, and symbolic regression \cite{vaddireddy2020feature,reinbold2021robust} have been adopted to help build/enrich the term library of the underlying model. Sparse regression methods, however, lack robustness due to issues including inaccurate numerical differentiation (especially with large noise) \cite{meidani2021data}, hard-thresholding \cite{zhang2021robust}, and incomplete library (even with library enriching techniques) \cite{xu2020dlga}. Therefore, a robust method of scientific discovery is necessitated for discovering the underlying physics from measured data that are usually incomplete and/or imprecise. 

Recently, fusing prior domain knowledge in solving computational physics problems has been widely recognized as potential and promising for improving the stability, generality, and robustness of the computational model. This potential has been demonstrated in physics informed machine learning \cite{raissi2019physics,wang2017physics,zhang2021structural} and hybrid modeling \cite{quaghebeur2021incorporating,rai2020driven}. The prior knowledge about the general physical principles has also been fused into discovering the governing PDE(s) when establishing the candidate model library \cite{reinbold2021robust,meidani2021data}. In ref. \cite{meidani2021data}, a machine learning classification method was proposed to detect the presence of physical patterns (such as diffusion and convection) in the investigated system and subsequently identify the governing PDE(s). Features used for this classification problem are extracted from the spatial and temporal derivatives of data and their statistics and frequency domain characteristics, considering the behaviors of physical patterns potentially existing in the investigated system. A classifier is first trained using a library of candidate models and then tested on the target system. Results show that the robustness of derived features is very limited when the level of noise goes above 0.1\%, despite the novelty of this scientific discovery scheme. 

In this study, we use the Euler characteristics (ECs) as features for characterizing dynamical systems and will examine its robustness when large noise exists in the measured/simulated data. EC is a general topological descriptor that characterizes the geometrical features of complex datasets with reduced dimensionality and complexity while preserving the maximum information \cite{SMITH2021107463}. Compared with other topological descriptors, EC has improved generality in characterizing a wide range of mathematical objects due to its fundamental connections with other descriptive tools such as statistics, field theory, and graph theory. Given a data object, its topology will be reduced to an EC curve by using the filtration transformation technique.

The general definition of the EC ($\chi$) for a 2D manifold is: $\chi$ $=$ $\beta_0$ $-$ $\beta_1$, in which $\beta_0$ is the number of 0-dimensional topological bases (known as connected components), $\beta_1$ is the number of 1-dimensional topological bases (known as holes), and $\beta_0, \beta_1 \in \mathbb{Z}_+$ where $\mathbb{Z}_+$ is the set of nonnegative integers. The relationship between the number of topological bases and dimensions can be extended to shapes with higher dimensions. The EC of an $(n+1)$-dimensional shape is defined as $\chi = \sum_{i=0}^n(-1)^i\beta_i$, in which $\beta_n \in \mathbb{Z}_+$ is the $n^\mathrm{th}$ Betti number and denotes the number of unique $n$-dimensional topological bases for the given shape \cite{adler2008some}.

\begin{figure}[ht]
\centering
\includegraphics[scale=1.0]{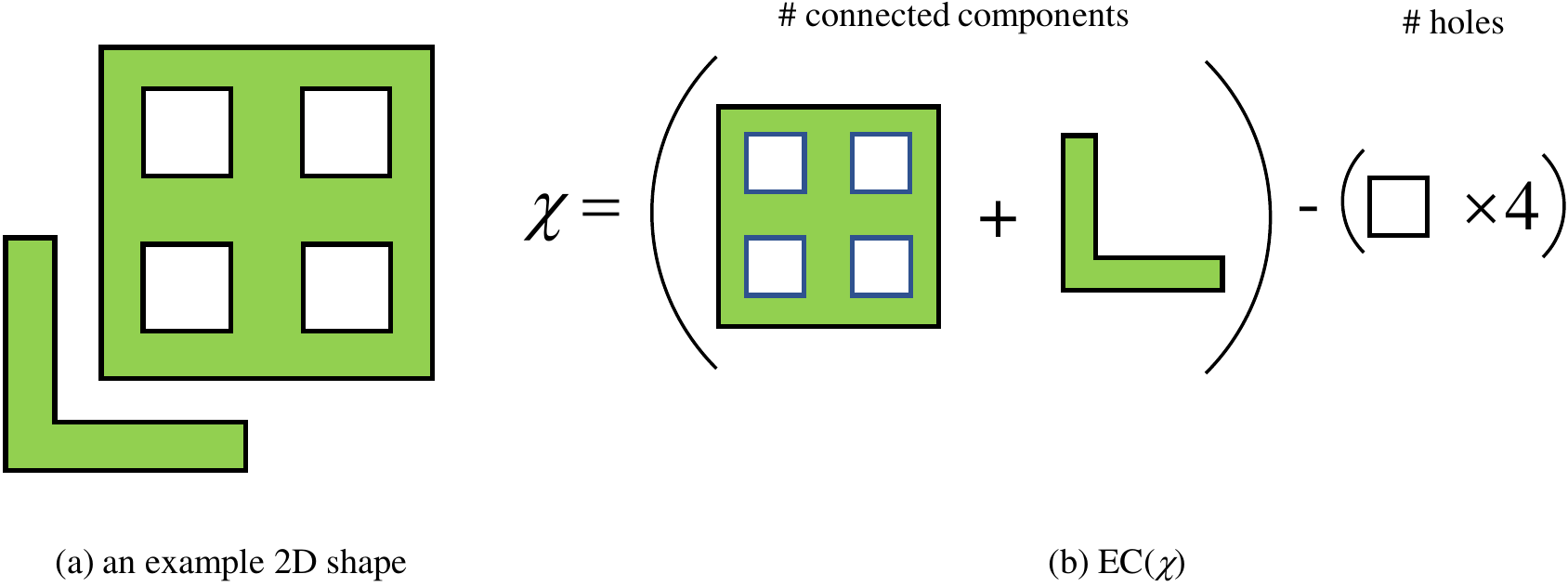}
\caption{Illustration on the calculation of EC for a 2D manifold. (a) A 2D shape with two connected components (i.e., $\beta_0=2$) and four holes (i.e., $\beta_1=4$). (b) The EC is an alternating sum of the number of connected components and holes (i.e., $\chi=-2$).}
\label{fig:EC2D}
\end{figure}

\begin{figure}[!ht]
\centering
\includegraphics[width=\linewidth]{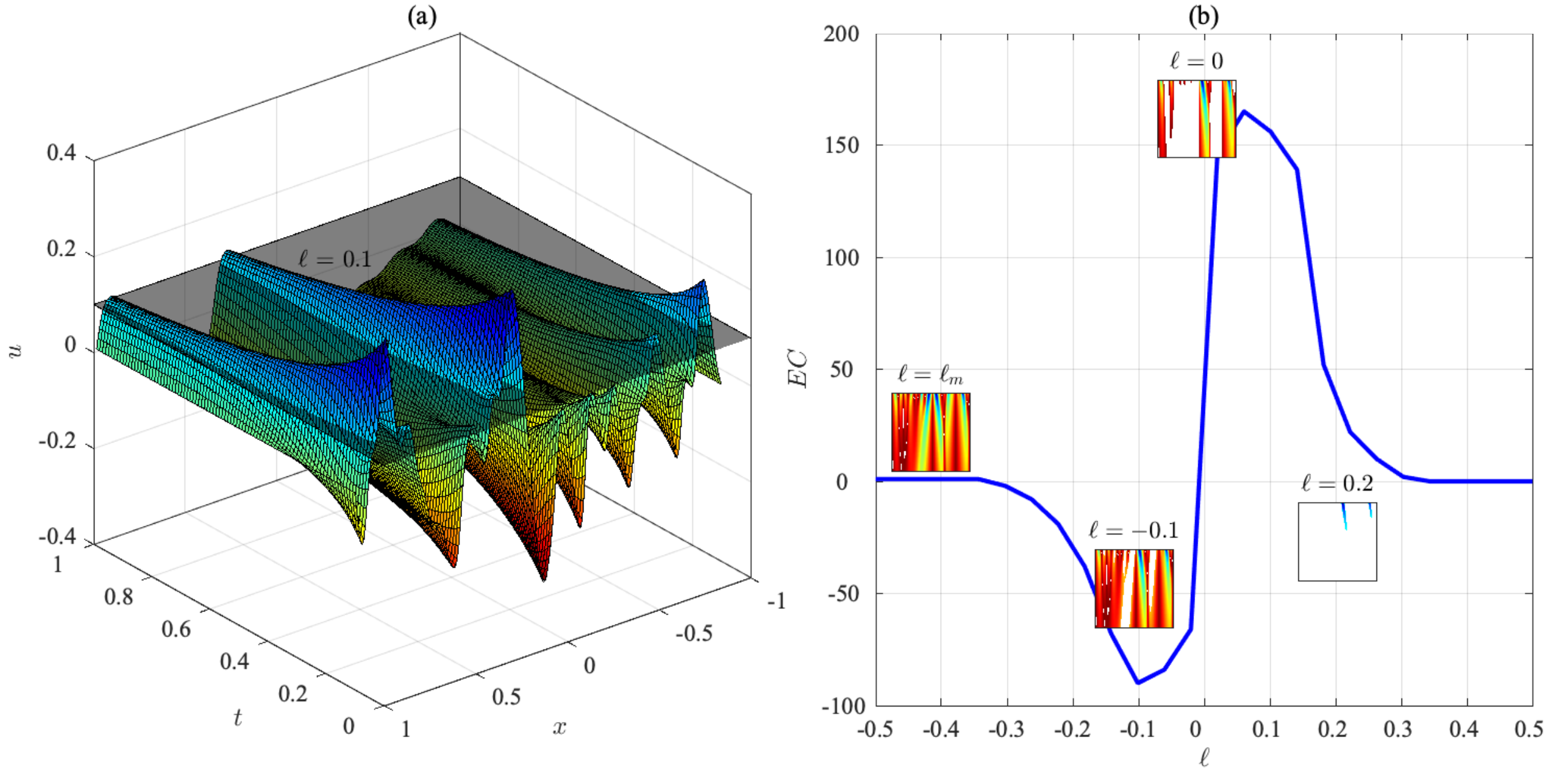}
\caption{(a) Superlevel set filtration of the 2D field (embedded in a 3D manifold) created with the solutions of the 1D Burgers equation. The plane surface represents the threshold of the filtration (i.e. $\ell$) cutting through the 3D field graph. As $\ell$ shifts from the top of the manifold to its bottom, the EC of each resulting field graph $G_\ell$ can be computed. (b) The EC curve constructed via filtration. This 2D curve captures the evolution of the topology of the varying superlevel sets during the filtration as shown in the four example snapshots.}
\label{fig:B1_EC_curve}
\end{figure}

For a dynamical system, its states (or solutions, i.e., $u$ or $\mathbf{u}$) lie in a high-dimensional field $\mathcal{D}_x \times \mathcal{D}_t$, in which $\mathcal{D}_x$ is the spatial domain and $\mathcal{D}_t$ is the temporal domain. As a result, an $n$-dimensional manifold $M$ will be formulated with the solutions using the chart $(X,f)$. Here $n$, the dimension of the topological space, is larger than the dimension of $\mathcal{D}_x$ by two due to the addition of the temporal coordinate and the function $f$; $X\subseteq \mathbb{R}^{n-1}$ is a Euclidean space formed by $X = \mathcal{D}_x \times \mathcal{D}_t$; the filed/function $f: X\rightarrow \mathbb{R}$ maps from $X$ to a scalar function of the system solutions. Taking the dissipative system characterized by the 1D Burgers equation $u_t = -uu_x+\nu u_{xx}$ ($\nu$ is the diffusion coefficient) as an example, its solution $u(x,t)$ lies in a 2D field and is embedded in a 3D manifold given by a 3D chart (i.e., $n=3$) (see Figure \ref{fig:B1_EC_curve} (a)). The function $f$ maps from the spatiotemporal coordinates (i.e., $x$ and $t$) to the scalar solutions (i.e., $u$) of the equation. The chart $(X,f)$ is also often referred to as the graph of field $f$. It should be noted that the graph does not live in a Euclidean space while $X$ does. As a result, the ECs and associated EC curve focus on the global topology of the graph during a filtration instead of the specific local connectivity information.

Filtration can be applied to manifolds by virtue of the concept of superlevel set \cite{poincare1895analysis}.  Given a manifold $M$ with field $f: X \rightarrow R$ and domain $X \subseteq \mathbb{R}^n$, the superlevel set $X_\ell$ at a threshold $\ell \in \mathbb{R}$ is defined as: $X_\ell=\{x\in X:f(x) \geqslant \ell$\}. The super level set has an associated field graph $G_\ell := G(X_\ell, f_\ell)$ with $f_\ell$ defined over $X_\ell$. The field graph $G_\ell = (X_\ell, f_\ell)$ contains all points of the manifold $M$ that have a function value greater than or equal to $\ell$ (it is a filtration of the manifold). The filtration creates a nested set of field graphs which are obtained by defining a sequence of decreasing filtration values $\ell_1 > \ell_2 > . . . > \ell_m$ with associated field graphs: $G_{\ell_1} \subseteq G_{\ell_2} \subseteq...\subseteq G_{\ell_m} \subseteq G$. Here, the field graphs are sparser with larger threshold values; we also have that $G_{\ell_m} = G(X,f)$ (the original graph) if $\ell_m = \mathrm{min}_{x\in X} f(x)$. For each superlevel set we obtain the field graph $G_\ell$ and we compute and record its EC value $\chi_\ell$ (e.g., we determine the number of connected components and number of holes). This information is used to construct an EC curve, which contains the pairs $(\ell, \chi_\ell)$. It is important to highlight that the EC curve provides a topological descriptor for a general $n$-dimensional field. The types of topological bases change with dimension.

\begin{figure}[!hb]
\centering
\includegraphics[scale=0.9]{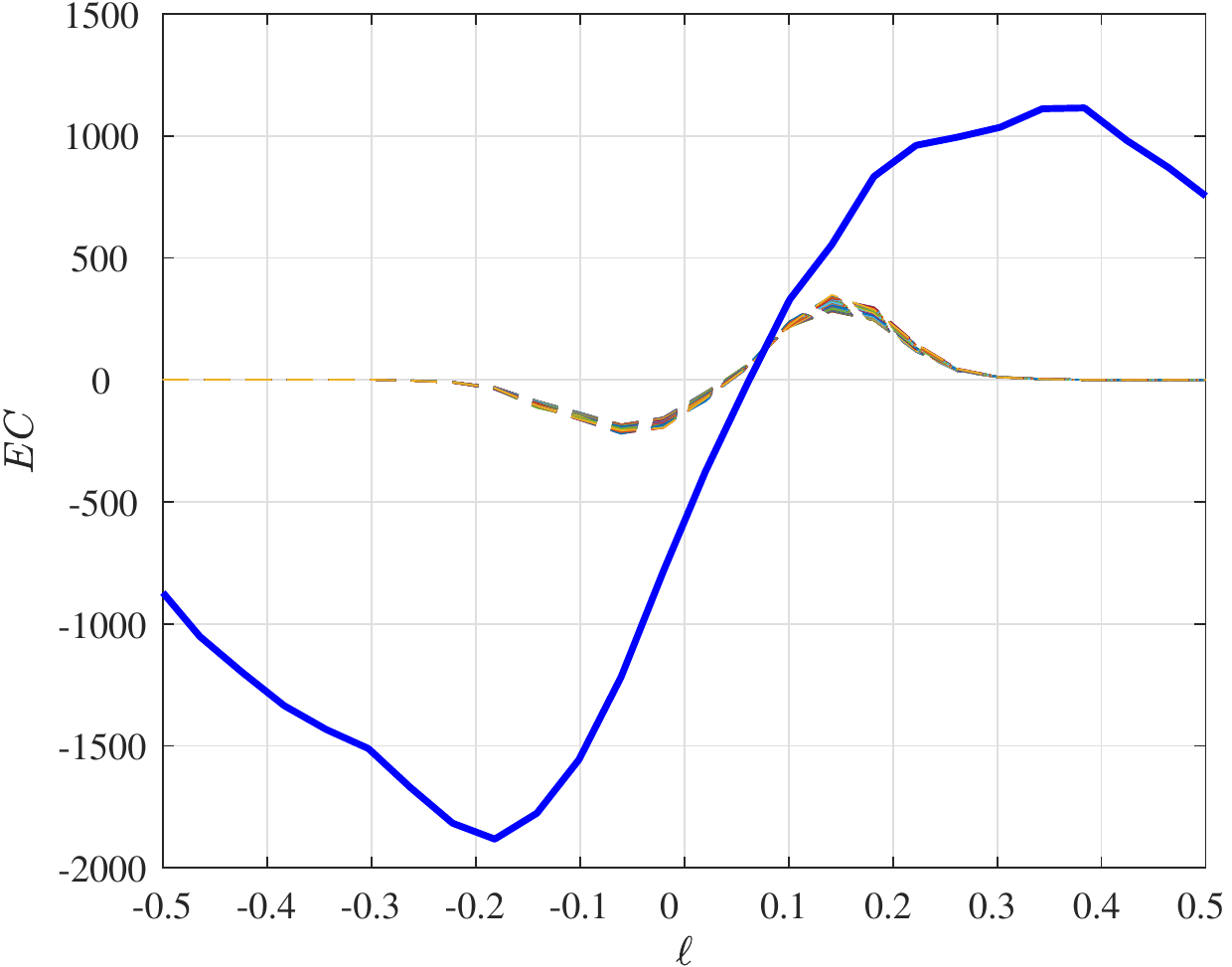}
\caption{EC curves created using noisy measurements from the 1D dissipative system. The dashed curves are EC curves from simulated data of an example case of the system  $u_t = \lambda_1uu_x+\lambda_2u_{xx}$ containing 0\% to 50\% noise. The solid curve is the EC curve for an example case of the system $u_t = \lambda_1uu_x+\lambda_2u^2u_{xx}$. We can observe that the difference of ECs caused by the added noise is much smaller than the that by the model difference.}
\label{fig:B1Noise}
\end{figure}

Taking the dissipative system characterized by the 1D Burgers equation as an example, with the solutions of the 1D Burgers equation that are embedded in a 3D manifold, the threshold in filtration is a 2D plane that cuts through the 3D graph yielding connected components and holes (see Figure \ref{fig:B1_EC_curve} (a)). When the plane passes through a local maximum we have that connected components are formed, when it passes a saddle point components are joined to form holes, and when a local minimum is passed holes are filled. This reveals that filtration captures the incidence of different types of critical points in the field (its topological features) and this information is summarized in the EC curve (see Figure \ref{fig:B1_EC_curve} (b)). Moreover, we found that the EC curve is immune from noise contamination (see Figure \ref{fig:B1Noise}), which has been mathematically proved in ref. \cite{ghrist2008barcodes}. This advantageous property of ECs over other non-topological features will largely improve the robustness of physics discovery especially when the measured data contain a large level of noise. Therefore, this methodology we propose in this study will potentially solve the most critical challenge in data-driven physics discovery \cite{rudy2017data,reinbold2021robust,zhang2021robust}. 

Given the measured data from an instrumented system, we formulate the manifold $M$ on which the filtration will be performed. The resulting EC vector will be used as features for characterizing the system and distinguishing it from others with a different (though maybe similar) governing equation. First, we establish a library of candidate models either using the results of sparse regression methods or with our knowledge about the investigated system. Data (i.e., $u$) and features (i.e. ECs) will be simulated with extensive combinations of random initial and boundary conditions so that the final decision regarding physics discovery will be as robust as possible. Simulated data may contain a certain level of measurement noise. We subsequently train a classifier using the simulated data with the label set as the ID of the system model that a dataset belongs to. Taking the above dissipative system as an example, with a collected noisy dataset, the $\Psi$-PDE method (see details in \textbf{Methods}) yields two candidate models: (1) $u_t = \lambda_1uu_x+\lambda_2u_{xx}$ and (2) $u_t = \lambda_1uu_x+\lambda_2u^2u_{xx}$. Then we simulate a collection of systems for each candidate model by assigning random values of $\lambda_1$ and $\lambda_2$ and specifying random initial conditions and boundary conditions. The outcome feature vectors from these two collections are labeled with the model IDs, i.e., 1 and 2, respectively. These simulated data are used to train a classifier. The Support Vector Machine (SVC) classifier is used in this study. The trained SVC model will be used to determine the governing model of a new system by feeding its output features into the classifier (see details of this methodology in \textbf{Methods}). To demonstrate the effectiveness of ECs in representing dynamical systems, the following scenarios are investigated in this study.
\begin{itemize}
\item First, we extend the analysis in ref. \cite{SMITH2021107463} for a reaction-diffusion system from the perspective of system identification and physics discovery. 
\item Second, we use this machine learning classification scheme to improve the confidence level of the sparse regression results. As mentioned above, most sparse regression methods lack robustness and may yield a considerably different model when the hyperparameter settings change slightly. The classification outcome using the EC features can help evaluate the representativeness of each outputted model. 
\item Third, we use the EC features to improve the accuracy of classification for identifying the 2D dynamical systems investigated in ref. \cite{meidani2021data}. The robustness will be examined when the data contain large measurement noise.
\end{itemize}

When investigating a 2D system, the $u$ component of the system states $\mathbf{u}$ is used for calculating the ECs, and we found that the identification results are not significantly affected when using the $v$ component for the EC calculations.

\section*{Results}
\subsection*{Charactering reaction and diffusion}
\begin{figure}[ht]
\centering
\includegraphics[width=\linewidth]{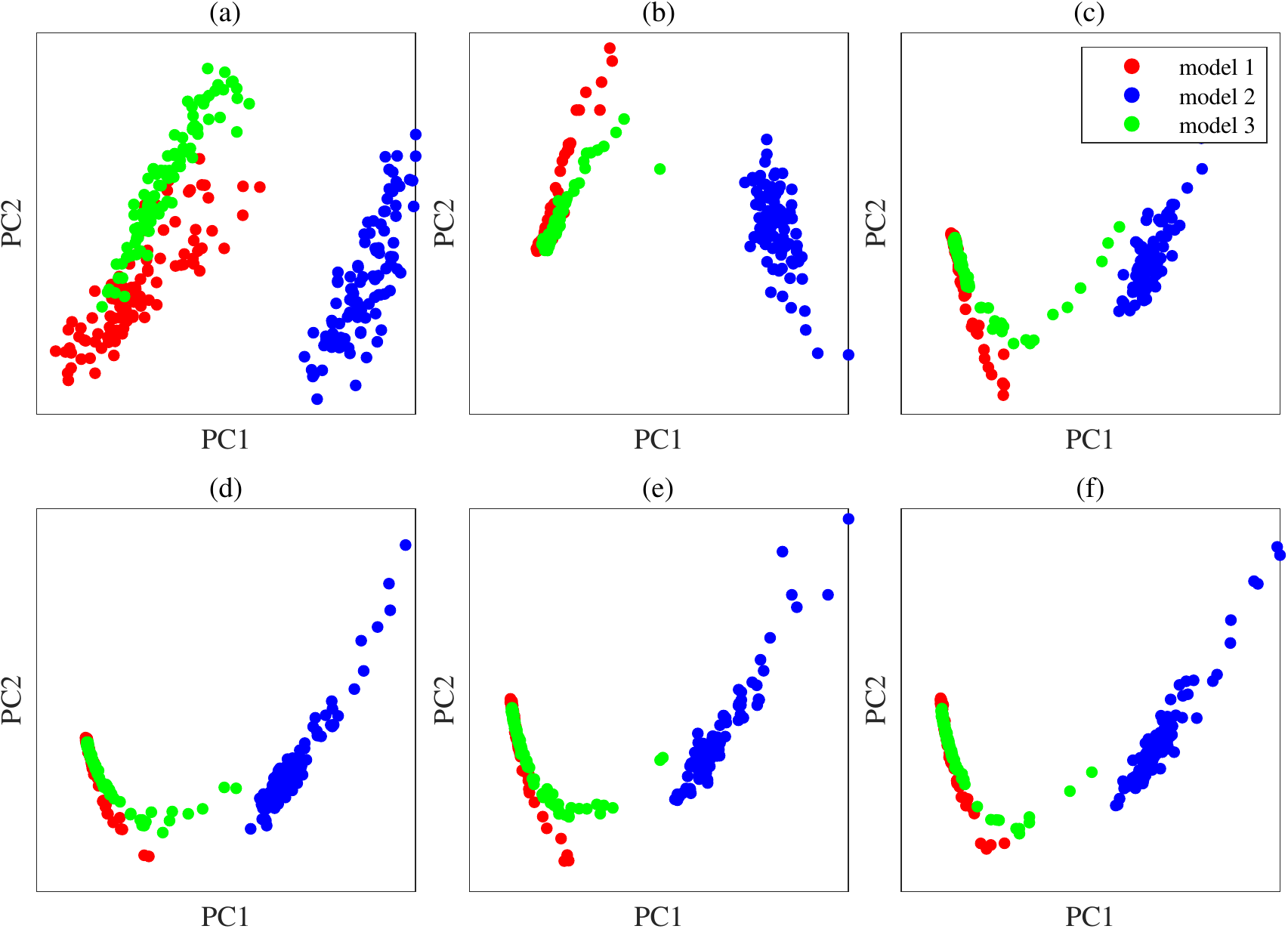}
\caption{SVD projection of the EC curves for the reaction-diffusion system onto their two leading principal components (PCs). Models 1, 2, and 3 denote the systems characterized by (1) $u_t = D(u_{xx}+u_{yy})+R(v-u)$ and $v_t = D(v_{xx}+v_{yy})+R(u-v)$, (2) $u_t = R(v-u)$ and $v_t = R(u-v)$, and (3)  $u_t = D(u_{xx}+u_{yy})$ and $v_t = D(v_{xx}+v_{yy})$, respectively.}
\label{fig:RD_PCA}
\end{figure}

\begin{figure}[ht]
\centering
\includegraphics[width=\linewidth]{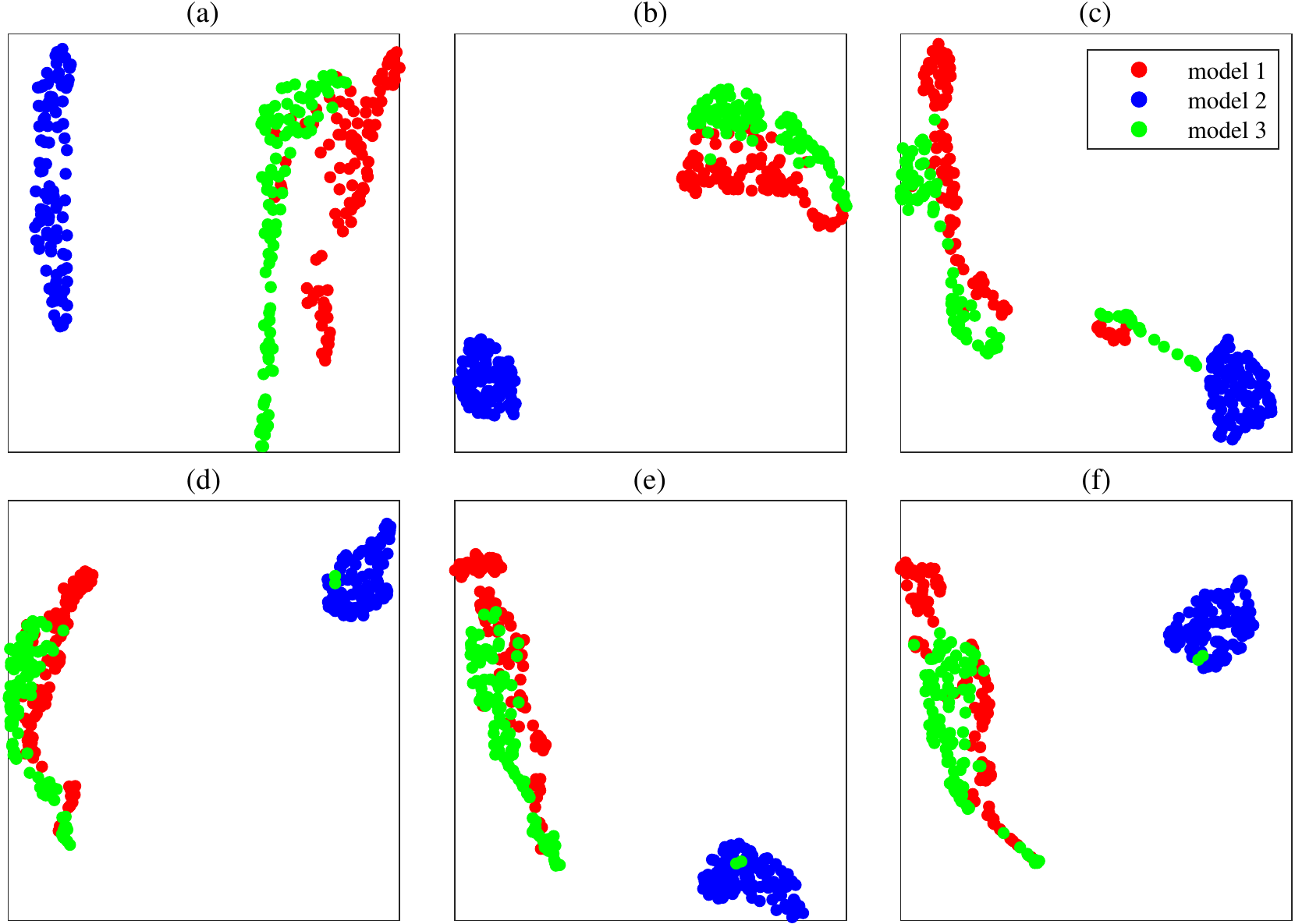}
\caption{UMAP embeddings of the EC curves for the reaction-diffusion system. Hyperparameter settings of UMAP are as follows: n\_neighbors = 15; min\_dist = 0.3; metric = `manhattan'.}
\label{fig:RD_Umap}
\end{figure}

In ref. \cite{SMITH2021107463}, ECs are used to characterize the 2D reaction-diffusion system ($u_t = D(u_{xx}+u_{yy})+R(v-u)$ and $v_t = D(v_{xx}+v_{yy})+R(u-v)$) with different nonzero reaction and diffusion coefficients (i.e., $R$ and $D$, respectively). In this study, we remove either the reaction or diffusion term from the governing equation such that we will have three candidate models in the library for the physics discovery problem: (1) $u_t = D(u_{xx}+u_{yy})+R(v-u)$ and $v_t = D(v_{xx}+v_{yy})+R(u-v)$, (2) $u_t = R(v-u)$ and $v_t = R(u-v)$, and (3)  $u_t = D(u_{xx}+u_{yy})$ and $v_t = D(v_{xx}+v_{yy})$. 100 examples are simulated for each model with random coefficient values and initial and boundary conditions (details about data generation can be found in \textbf{Methods}). From the reduced representations (see Figures \ref{fig:RD_PCA} and \ref{fig:RD_Umap}), we found that the three systems can be well separated using the ECs. Both the SVD projection (see Figure \ref{fig:RD_PCA}) and the UMAP embeddings (see Figure \ref{fig:RD_Umap}) show that examples of model (2) with only the reaction term are more ``distant'' from that of the other two models that both have the diffusion term. We speculate from this phenomenon that this 2D system is more dominated by the diffusion term than the reaction term. The average testing accuracy using the trained classifier is 0.95, 0.93, 0.98, 0.92, 0.97, and 0.98 for the cases with 0\%, 10\%, 20\%, 30\%, 40\%, 50\% measurement noise, respectively. Therefore, we found in this analysis that the ECs can be used for identifying the 2D reaction-diffusion system in addition to distinguishing the system with different model parameter values (as shown in ref. \cite{SMITH2021107463}).

\subsection*{Classification for confidence enhancement}
\begin{figure}[ht]
\centering
\includegraphics[width=\linewidth]{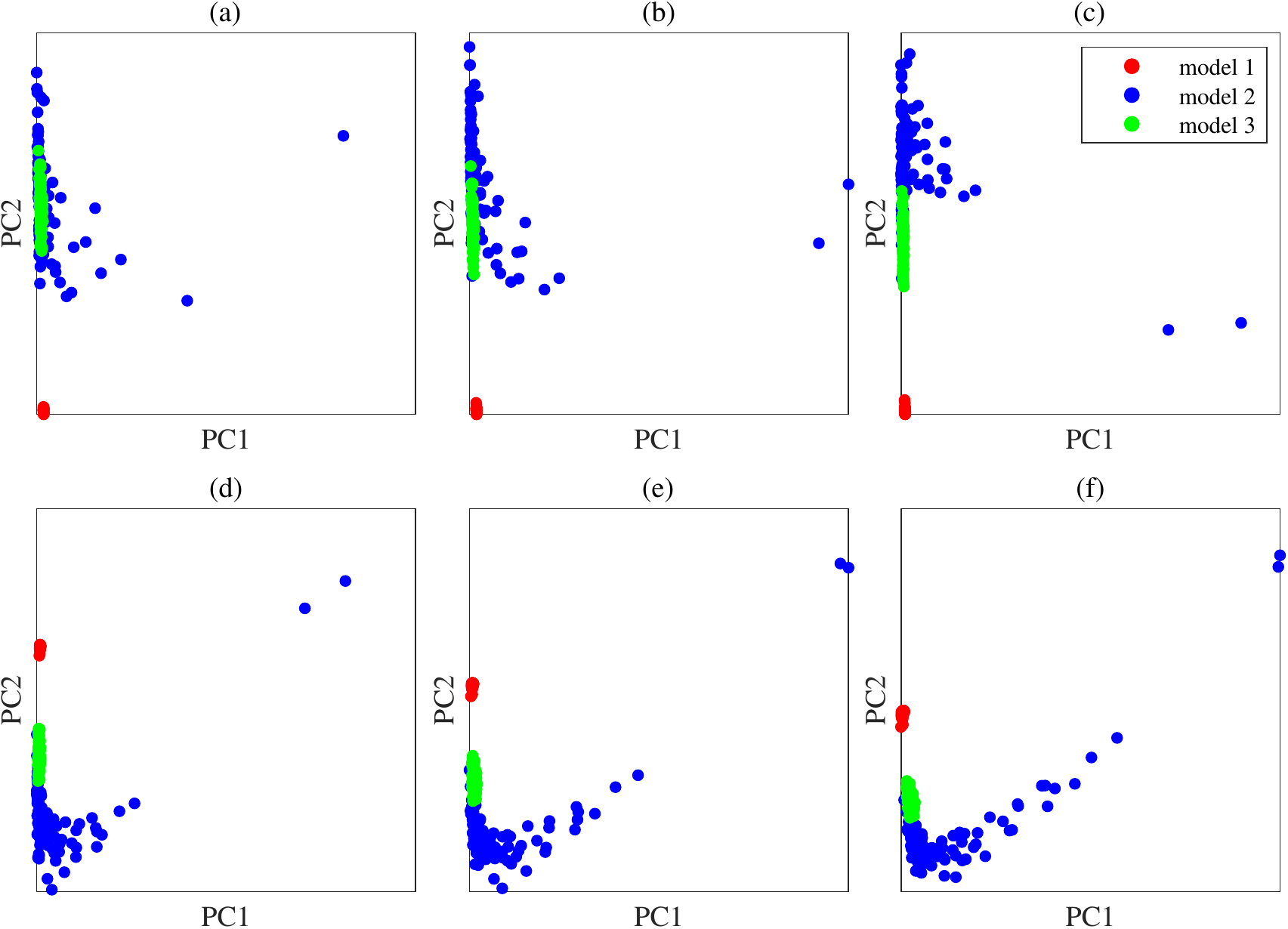}
\caption{SVD projection of the EC curves for the 1D dissipative system onto their two leading principal components (PCs). Models 1, 2, and 3 denote the systems characterized by (1) $u_t = \lambda_1uu_x + \lambda_2u_{xx}$, (2) $u_t = \lambda_1uu_x + \lambda_2u^2u_{xx}$, and (3) $u_t = \lambda_1u_{xx}$, respectively. (a) to (f) show the case with 0\% noise, 10\% noise, 20\% noise, 30\% noise, 40\% noise, and 50\% noise, respectively.}
\label{fig:B1PCA}
\end{figure}

\begin{figure}[ht]
\centering
\includegraphics[width=\linewidth]{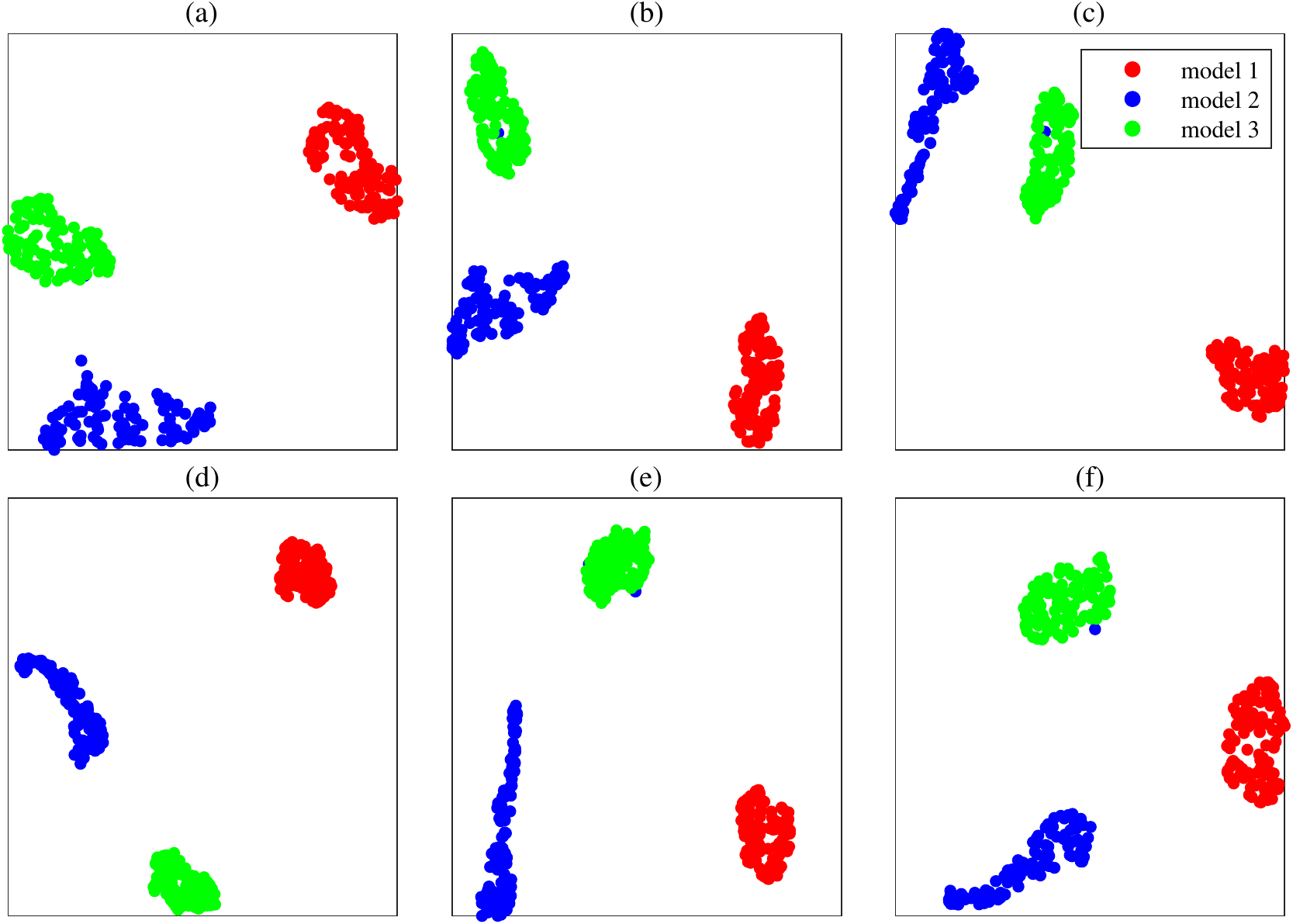}
\caption{UMAP embeddings of the EC curves for the 1D dissipative system. Hyperparameter settings of UMAP are as follows: n\_neighbors = 15; min\_dist = 0.3; metric = `manhattan'.}
\label{fig:B1Umap}
\end{figure}

In this section, we use the classification scheme to enhance the outcome of sparse regression methods of physics discovery. The 1D dissipative system characterized by the Burgers equation is used as an example. With the simulated data from this system, the $\Psi$-PDE method recommends two candidate models: (1) $u_t = \lambda_1uu_x+\lambda_2u_{xx}$ and (2) $u_t = \lambda_1uu_x+\lambda_2u^2u_{xx}$. Our previous study \cite{zhang2021robust} shows that it could be considerably challenging to make a final decision between these two models if the initial and boundary conditions are not known a priori. In addition to the above two models from the $\Psi$-PDE method, we added a third model into the model library: (3) $u_t = \lambda_1uu_x$ for better comparison. We found that the simulated examples from the three models can be well separated by using the ECs as characterizing features (see Figures \ref{fig:B1PCA} and \ref{fig:B1Umap}). Moreover, the measurement noise (as much as 50\%) does not undermine the representativeness of ECs in characterizing the system models. As expected, the average testing accuracy using the trained classifier is satisfactorily high (0.99, 0.99, 0.99, 1.00, 1.00, and 1.00 for cases with 0\% to 50\% noise, respectively). Comparing the results with previous studies in the literature, we can find that our methods outperform many existing methods of physics discovery including the popular PDE-FIND method \cite{rudy2017data} and DLGA-PDE method \cite{xu2020dlga}, in terms of its robustness when large measurement noise exists in data.

\subsection*{Identification of 2D system in ref. \cite{meidani2021data}}
\begin{figure}[ht]
\centering
\includegraphics[width=\linewidth]{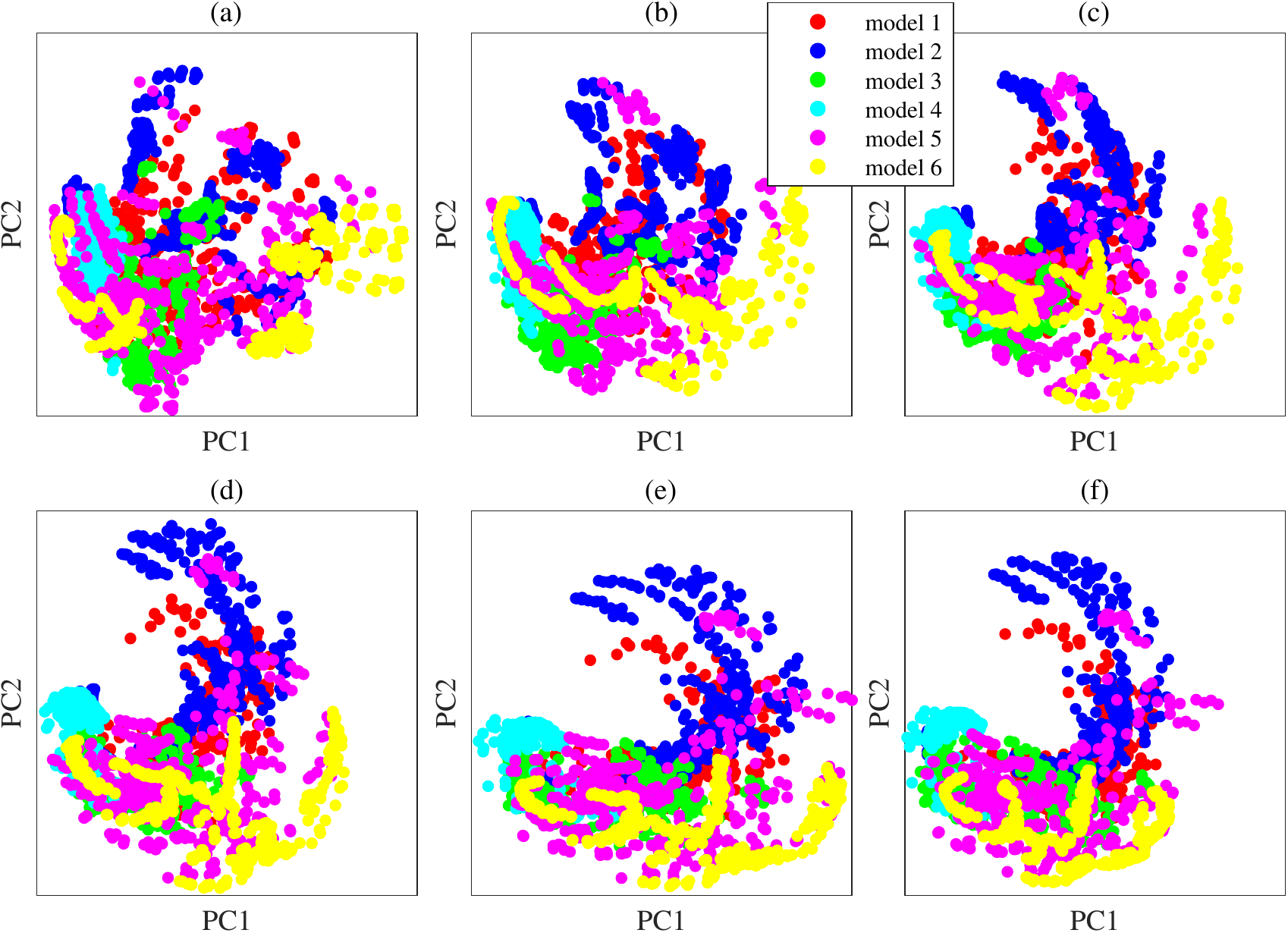}
\caption{SVD projection of the EC curves for the 2D systems in Table \ref{tab:PDE2D_SVC} onto their two leading principal components (PCs). (a) to (f) show the case with 0\% noise, 10\% noise, 20\% noise, 30\% noise, 40\% noise, and 50\% noise, respectively.}
\label{fig:2D_PCA}
\end{figure}

\begin{figure}[ht]
\centering
\includegraphics[width=\linewidth]{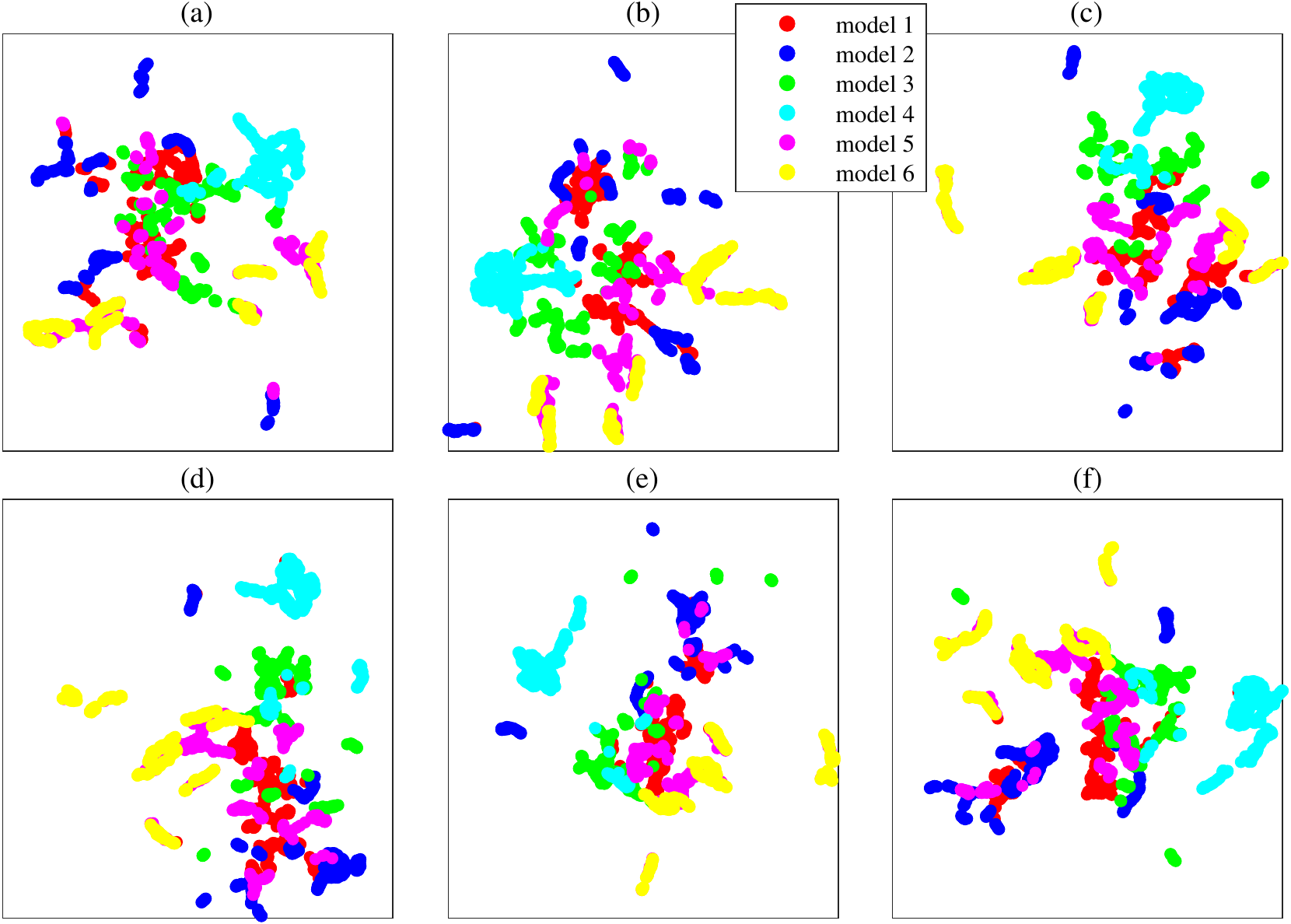}
\caption{UMAP embeddings of the EC curves for the 2D systems in Table \ref{tab:PDE2D_SVC}. Hyperparameter settings of UMAP are as follows: n\_neighbors = 15; min\_dist = 0.3; metric = `manhattan'.}
\label{fig:2D_Umap}
\end{figure}

In this section, we use the EC features to characterize the six 2D spatiotemporal systems in ref. \cite{meidani2021data} (see column 2 of Table \ref{tab:PDE2D_SVC} for the governing equations). It can be observed that many models share one or more terms in common, which significantly amplifies the challenge of data-driven system identification. As a result, these models are not as well separated as in the above studies (see Figures \ref{fig:2D_PCA} and \ref{fig:2D_Umap}). It should be noted that both the SVD projection and the UMAP embeddings are reduced representations of the original data and may not fully reflect the characterizing capability of the ECs. The classification results in Table \ref{tab:PDE2D_SVC} show that we can accurately identify the system models using the ECs as features. Compared with the hand-crafted features in  ref. \cite{meidani2021data} (in which the average accuracy reduces from 98.7\% to 86.4\% when 0.5\% noise is added to the clean data), the ECs are much more robust in cases with noisy data. In this section, we demonstrate that the EC-based classification method can be used for identifying a wide range of 2D dynamical systems.

\begin{table}[ht]
\centering
\begin{tabular}{|l|l|l|l|l|l|l|l|}\hline
\multirow{2}{*}{model class} &\multirow{2}{*}{model equation} & \multicolumn{2}{l|}{precision} & \multicolumn{2}{l|}{recall} &  \multicolumn{2}{l|}{f1-score} \\ \cline{3-8}
& & 0\% noise & 50\% noise & 0\% noise & 50\% noise & 0\% noise & 50\% noise \\ \hline
1& $u_{t} - c \nabla^{2}u  = 0$ & 0.96 & 0.98 & 0.97 & 0.94 & 0.97 & 0.96 \\ \hline
2& $u_{t} - c \nabla^{2}u + B\nabla u = 0$ & 0.99 & 0.98 & 0.99 & 0.97 & 0.99 & 0.98\\ \hline
3& $u_{tt}  - c \nabla^{2}u  = 0$ & 0.98 & 0.96 & 0.96 & 0.91 & 0.97 & 0.93 \\ \hline
4& $u_{tt} - c \nabla^{2}u + B\nabla u = 0$ & 0.96 & 0.98 & 0.99 & 0.95 & 0.98 & 0.96 \\ \hline
5& $u_{tt} + du_{t} - c \nabla^{2}u = 0$ & 0.99 & 0.84 & 0.89 & 0.91 & 0.94 & 0.87 \\ \hline
6& $u_{tt} + du_{t} - c \nabla^{2}u + B\nabla u = 0$ & 0.94 & 0.93 & 1.00 & 1.00 & 0.97 & 0.96 \\ \hline
\end{tabular}
\caption{\label{tab:PDE2D_SVC}The classification results for the 2D systems. The average classification accuracy is 0.97 and 0.95 for the cases with 0\% and 50\% noise, respectively.}
\end{table}

\section*{Discussion}
As we have demonstrated in this study, the ECs can be used as representative features for characterizing a dynamical system and distinguishing it from other systems with different governing models. The robustness of the data-driven physics discovery using ECs has been verified with data containing a wide range of levels of measurement noise. Compared with previous sparse regression methods, the proposed method does not require implementing the fragile numerical differentiation, non-stable hyperparameter tuning, and hard-thresholding that requires empirical specification of the threshold values. Moreover, in this study, we provide an alternative (i.e., machine learning classification) to make confident choices among the candidate models provided by sparse regression. The adoption of EC features largely improves the robustness of the classification scheme proposed in ref. \cite{meidani2021data} and makes this scheme more feasible for practical application. Additionally, the method we proposed in this study requires much less data than most existing methods of physics discovery. The calculation of topological features (such as ECs) does not require a dense grid for the solutions of the system (i.e. $u$ or $\mathbf{u}$), unlike the numerical differentiation in sparse regression methods (such as ref. \cite{rudy2017data}) for preparing the library of candidate terms or in classification methods (such as in ref. \cite{meidani2021data}) for calculating the spatial/temporal derivative based features. This advantage can largely reduce the amount of data for discovering the underlying physics of a new dynamical system.

Although the results validate the potential of data-driven physics discovery using the EC-boosted classification approach, it requires establishing a complete (or overcomplete) model library to make the decision about the governing model. In case establishing such a library is not guaranteed, especially for a novel system for which little prior knowledge is available, a purely data-driven approach can be used to compensate for the discovered physics model. Hybrid system modeling integrates first-principle physics modeling with a data-driven approach (such as a black-box neural network model) for improved accuracy and robustness of system modeling. Our method can be used to enhance the confidence level of the physics modeling part so that the black-box modeling part has less burden of correcting the misrepresented physics.

\section*{Methods}\label{Sec:Method}
\subsection*{Experimental system and data collection.}
One 1D system and seven 2D systems (including the reaction-diffusion system in ref. \cite{SMITH2021107463}  and six 2D systems in ref. \cite{meidani2021data}) are investigated in this study.The 1D dissipative system is characterized by the Burgers equation $u_t = -uu_x+\nu u_{xx}$ with $\nu$ being the diffusion coefficient. The range of its solution is specified as $t\in [0,1]$ and $x\in [-1,1]$. The time and spatial ranges are discretized into 99 and 255 uniform intervals, respectively. As a result, $\delta t = 1/99$, $\delta x = 2/255$, and the system state $u$ has a dimension of $100\times 256$. Random initial and boundary conditions are adopted to demonstrate the generality of the method. 0\% to 50\% white Gaussian noise is added to the synthetic data to examine the robustness of the method when a considerable level of noise exists. In this study, the noise level is quantified by the percentage of the standard deviation of the measured variable. For example, if 10\% noise is added to the solutions $u$, then the outcome is $u_n = u+10\%*std(u)*randn(size(u))$ where $randn(\cdot)$ generates white Gaussian noise of the specified dimension. The code for generating the solutions of this system can be found in the \textbf{Data availability} and \textbf{Code availability} sections. The data for the 2D reaction-diffusion system investigated in this study are the same with that used in ref. \cite{SMITH2021107463} and can be found in \textbf{Data availability}. The data for the rest six 2D systems investigated in this study (listed in Table \ref{tab:PDE2D_SVC} are the same with that used in ref. \cite{meidani2021data} and can be found in \textbf{Data availability}.)

\subsection*{$\Psi$-PDE}
The $\Psi$-PDE method proposed in the authors' previous study \cite{zhang2021robust} is used to generate candidate models for the investigated 1D dissipative system. First, the collected data are split into training and validation sets and processed using a neural network model with the early-stopping strategy. This step is expected to significantly remove the measurement noise. Following the data preprocessing, the spatial and temporal derivatives are numerically calculated using the finite difference or polynomial interpolation method, and the library matrix will be formulated with the spatial derivatives. Subsequently, the fast Fourier transform (FFT) will be implemented to transform the sparse regression problem to the frequency domain and further reduce the influence of noise via frequency cutoff. Finally, the $\psi$-algorithm yields several candidate models. More details of the $\Psi$-PDE method can be found in ref. \cite{zhang2021robust}.

\subsection*{Robust physics discovery via classification using the EC features}
In this study, the machine learning classification method is applied to examine and subsequently improve the confidence in the learned model terms. With the spatiotemporal data collected from the investigated system, the collection of candidate models can be built whether using the sparse regression results (as in ref. \cite{zhang2021robust}) or with the prior knowledge about the system (as in ref. \cite{meidani2021data}). Subsequently, the data are simulated for all the candidate models with random settings of model parameter values, initial and boundary conditions, and the noise level. Following this step, ECs are calculated for each dataset and labeled with the model ID that the dataset is generated from. Next, with the labeled data, an SVM classifier can be trained with the hyperparameter tuned using a grid search strategy. Finally, the measured data from the investigated system are fed into the trained classifier for testing, and the model yielding the highest classification score is determined as the most probable model for this dynamical system. 

\begin{figure}[!ht]
\centering
\includegraphics[scale=0.6]{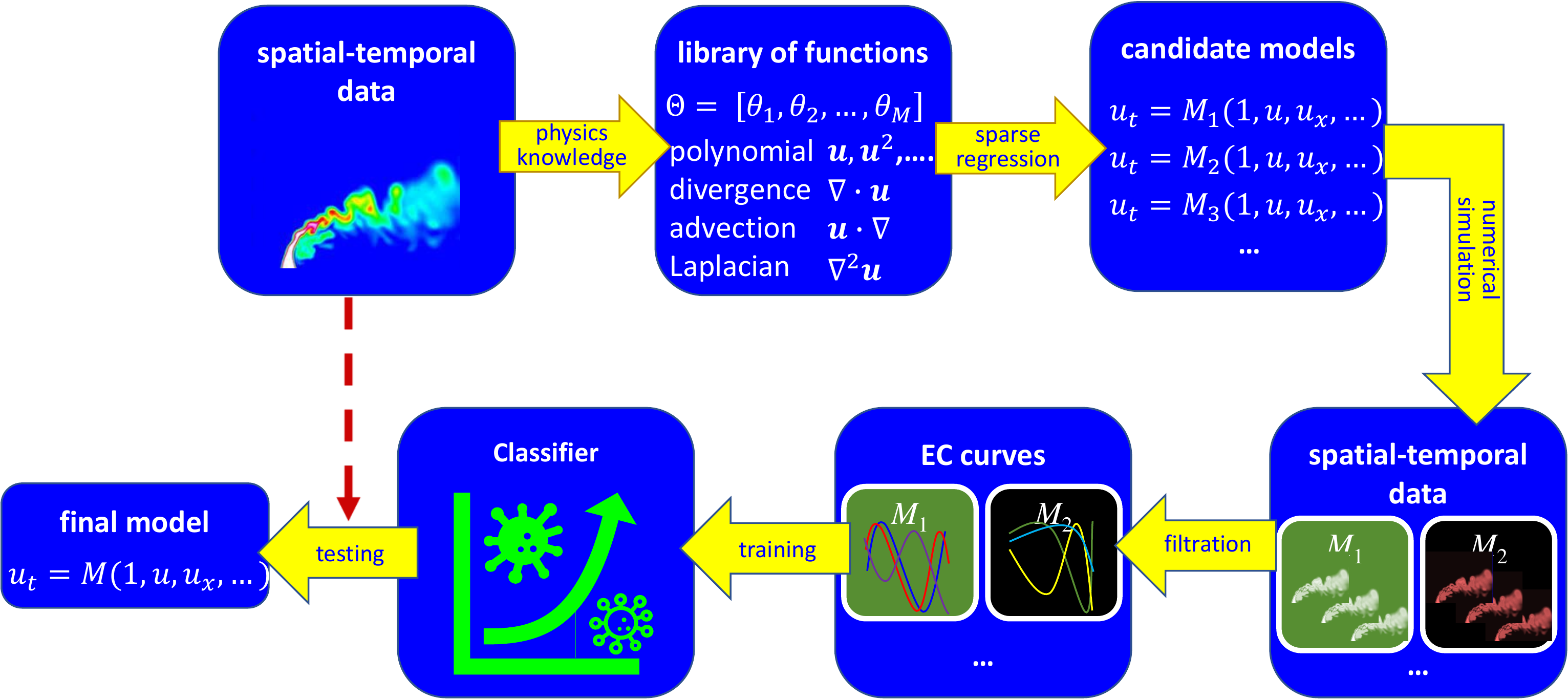}
\caption{The framework of robust physics discovery via classification using ECs. (1) This framework starts from the spatiotemporal data collected from the investigated dynamical system. (2) For sparse regression, the library of candidate terms is built with the physics knowledge. (3) Spare regression is implemented using the $\Psi$-PDE method, which usually yields more than one candidate model. (4) Solutions of all candidate models are simulated with an extensive combination of model parameters and initial and boundary conditions. (5) Filtration is implemented to generate EC curves for each simulated dataset, and each EC vector is labeled with the corresponding model ID. (6) An SVM classifier is trained using the labeled data. (7) The most representative model is determined by feeding the original dataset into the trained classifier.}
\label{fig:EC_classify}
\end{figure}

Taking the dissipative system characterized by the Burgers equation $u_t = -uu_x+0.01u_{xx}$ as an example, the $\Psi$-PDE method yields two candidate models: (1) $u_t = \lambda_1uu_x+\lambda_2u_{xx}$ and (2) $u_t = \lambda_1uu_x+\lambda_2u^2u_{xx}$, which constitute the collection of candidate models together with a third candidate model: (3) $u_t = \lambda_1uu_x$. Then the spatiotemporal data and corresponding ECs are simulated for the three models with different parameter combinations. Finally, the trained classifier yields model (1) when tested on the original dataset. Then we decide that the governing equation for this system is $u_t = \lambda_1uu_x+\lambda_2u_{xx}$. The framework of this method is shown in Figure \ref{fig:EC_classify}.

\section*{Data availability}

The data used for 1D Burgers equation and 2D reaction-diffusion equation can be found on GitHub at https://github.com/ymlasu/EC-PDE. The data used for 2D system identification in Table \ref{tab:PDE2D_SVC} can be found on Github at https://github.com/BaratiLab/PDE-Identification-Features referring to ref. \cite{meidani2021data}.

\section*{Code availability}

All python codes used for physics discovery in this study can be found on GitHub at https://github.com/ymlasu/EC-PDE. Any other requests should be made to the corresponding author.

\bibliography{EC_PDE}

\section*{Acknowledgements}

The research reported in this paper was supported by funds from NASA University Leadership Initiative Program (Contract No. NNX17AJ86A, Project Officer: Dr. Anupa Bajwa, Principal Investigator: Dr. Yongming Liu). The support is gratefully acknowledged.

\section*{Author contributions statement}

Z.Z. was responsible for the concept, research design, data analysis, and result interpretation. N.X. provided assistance in the manifold learning using the UMAP method. Y.L. supervised this research and was involved in the conceptualization and research design. All authors were involved in the manuscript writing and correction.

\section*{Competing interests}

The authors declare no competing interests.

\section*{Additional information}

\end{document}